\documentclass[manuscript]{aastex}
\def\lb#1{{\protect\linebreak[#1]}}

\def\httppisa{{\tt http://\lb{2}newton.\lb{2}dm.\lb{2}unipi.\lb{2}it/orbfit/}}

\slugcomment{Prepared for AJ}
\shorttitle{Spin vector and shape of (6070) Rheinland ...}
\shortauthors{}

\begin{document}

\title{Spin vector and shape of (6070) Rheinland and their implications}

\author{David Vokrouhlick\'y}
\affil{Institute of Astronomy, Charles University,
       V Hole\v{s}ovi\v{c}k\'ach 2, CZ--18000 Prague 8, \\
       Czech Republic, E-mail: vokrouhl@cesnet.cz}
\author{Josef \v{D}urech}
\affil{Institute of Astronomy, Charles University,
       V Hole\v{s}ovi\v{c}k\'ach 2, CZ--18000 Prague 8, \\ Czech Republic}
\author{David Polishook}
\affil{Department of Particle Physics and Astrophysics, Weizmann Institute
       of Science, Rehovot 76100, Israel}
\author{Yurij N. Krugly}
\affil{Institute of Astronomy, Karazin Kharkiv National University, Sumska 35,
       Kharkiv 61022, Ukraine}
\author{Ninel N. Gaftonyuk}
\affil{Crimean Astrophysical Observatory, Simeiz Department, Simeiz 98680,
       Ukraine}
\author{Otabek A. Burkhonov}
\affil{Ulugh Beg Astronomical Institute, Uzbek Academy of Sciences,
       Astronomicheskaya 33, Tashkent 100052, Uzbekistan}
\author{Shukhrat A. Ehgamberdiev}
\affil{Ulugh Beg Astronomical Institute, Uzbek Academy of Sciences,
       Astronomicheskaya 33, Tashkent 100052, Uzbekistan}
\author{Rivkat Karimov}
\affil{Ulugh Beg Astronomical Institute, Uzbek Academy of Sciences,
       Astronomicheskaya 33, Tashkent 100052, Uzbekistan}
\author{Igor E. Molotov}
\affil{Keldysh Institute of Applied Mathematics, RAS, Miusskaya
       sq.~4, Moscow 125047, Russia}
\author{Petr Pravec}
\affil{Astronomical Institute, Czech Academy of Sciences, Fri\v{c}ova 298,
       CZ--251 65 Ond\v{r}ejov, Czech Republic}
\author{Kamil Hornoch}
\affil{Astronomical Institute, Czech Academy of Sciences, Fri\v{c}ova 298,
       CZ--251 65 Ond\v{r}ejov, Czech Republic}
\author{Peter Ku\v{s}nir\'ak}
\affil{Astronomical Institute, Czech Academy of Sciences, Fri\v{c}ova 298,
       CZ--251 65 Ond\v{r}ejov, Czech Republic}
\author{Julian Oey}
\affil{Leura Observatory, 94 Rawson Pde., Leura, NSW, Australia}
\author{Adri\'an Gal\'ad}
\affil{Modra Observatory, Comenius University, Bratislava SK-84248, Slovakia}
\author{Jind\v{r}ich \v{Z}i\v{z}ka}
\affil{Institute of Astronomy, Charles University,
       V Hole\v{s}ovi\v{c}k\'ach 2, CZ--18000 Prague 8, \\ Czech Republic}

\begin{abstract}
 Main belt asteroids (6070) Rheinland and (54827) 2001~NQ8 belong to
 a small population of couples of bodies which reside on very similar
 heliocentric orbits. Vokrouhlick\'y \& Nesvorn\'y (2008, AJ 136, 280)
 promoted a term ``asteroid pairs'', pointing out their common origin within
 the past tens to hundreds of ky. Previous attempts to reconstruct the
 initial configuration of Rheinland and 2001~NQ8 at the time of their
 separation have led to the prediction that Rheinland's rotation should be
 retrograde. Here we report extensive photometric observations of this
 asteroid and use the lightcurve inversion technique to directly determine
 its rotation state and shape. We confirm the retrograde sense of 
 rotation of Rheinland, with obliquity value constrained to be
 $\geq 140^\circ$. The ecliptic longitude of the pole position is not
 well constrained as yet. The asymmetric behavior of Rheinland's
 lightcurve reflects a sharp, near-planar edge in our convex shape
 representation of this asteroid. Our calibrated observations in the red
 filter also allow us to determine $H_R = 13.68\pm 0.05$ and $G =
 0.31\pm 0.05$ values of the H-G system. With the characteristic color
 index $V-R = 0.49\pm 0.05$ for the S-type asteroids, we thus obtain $H =
 14.17\pm 0.07$ for the absolute magnitude of (6070) Rheinland. This a
 significantly larger value than previously obtained from analysis of the
 astrometric survey observations. We next use the obliquity constraint for 
 Rheinland to eliminate some degree of uncertainty in the past
 propagation of its orbit. This is because the sign of the past secular
 change of its semimajor axis due to the Yarkovsky effect is now
 constrained. Determination of the rotation state of the secondary
 component, asteroid (54827) 2001~NQ8, is the key element in further
 constraining the age of the pair and its formation process.
\end{abstract}


\section{Introduction}
Pairs of asteroids residing on very similar heliocentric orbits were
recently discovered in the Hungaria population and in the main belt (e.g.,
Vokrouhlick\'y \& Nesvorn\'y 2008; Pravec \& Vokrouhlick\'y 2009;
Milani et~al. 2010). The orbits  of components in a pair, often too similar
to be a random fluke in the background population of asteroids, suggests 
a common origin. Indeed, by backward integration of orbits of paired asteroids,
we were able to identify, for most cases, specific epochs in the past 
tens to hundreds of kys when the two components become very close to 
each other. These close encounters were interpreted as formation events 
of the pairs during which the two components gently separated from a 
common parent body.

Asteroid pairs thus share some fundamental properties with the related
asteroid families, the similarity being the most apparent for the very 
young families (e.g., Nesvorn\'y et~al. 2006; Nesvorn\'y \& Vokrouhlick\'y
2006; Vokrouhlick\'y \& Nesvorn\'y 2011): notably, members in both 
pairs and families arise as fragments from a disintegrated parent asteroid.
However, it has been unclear whether they also share a common
formation process. Indeed, while the larger asteroid families are obviously
of collisional origin, Vokrouhlick\'y \& Nesvorn\'y (2008) have
discussed several other putative formation processes for the asteroid
pairs. The hunt for identification of the formation process of the
asteroid pairs motivated Pravec et~al. (2010) to conduct photometric
observations of the primary (larger) components in numerous pairs.
Their main results can be summarized as follows: (i) there is a strong 
correlation between the rotation period of the primary component and 
mass ratio of the two asteroids in the pair, and (ii) there is a lack
of pairs with mass ratio of the two asteroids larger than $\simeq 0.2$. 
The asymptotic behavior of (i) above is as follows: in pairs where 
one component is much smaller than the other, the primaries 
systematically rotate very fast (near the rotation fission barrier 
observed for solitary asteroids; e.g., Pravec et~al. 2002), whereas in 
pairs that have a smaller mass ratio between the larger and smaller 
components, the primaries
systematically rotate very slow. These observations convincingly
demonstrate that most of the asteroid pairs were formed by rotational
fission rather than catastrophic (collisional) breakup of the parent
body (cf. Pravec et~al. 2010). The YORP effect%
\footnote{The YORP effect is due to torques of scattered
 sunlight on the asteroid surface, as well as those due to the thermal
 radiation of the body itself (e.g., Bottke et~al. 2002, 2006).}
has been suggested as the underlying physical mechanism that brought 
the parent body rotation to the fission limit.

To further characterize the principal formation process of the asteroid
pairs, it is important to both (i) continue observations of parameters
of the whole population, and (ii) also characterize selected pairs as
precisely as possible. This work goes along the (ii) line. Already
Vokrouhlick\'y \& Nesvorn\'y (2008) recognized that the pair of
asteroids (6070) Rheinland and (54827) 2001~NQ8 is somewhat exceptional
among other known pairs since it allows the most precise determination
of its age. This is because the age is young, $\simeq 17$~kyr only,
and the two asteroids are large enough such that effects of both dynamical
chaos and thermal forces are minimized in their past orbital evolution. 
Vokrouhlick\'y \& Nesvorn\'y 
(2009) extended and substantiated the previous work by taking into
account also mutual gravitational forces of the two components in the
initial phase of their separation. Statistical analysis of the angle
between the angular momenta of the heliocentric orbital motion of
Rheinland and the mutual motion of the two components at their
separation let these authors to conjecture that Rheinland's rotation
should be preferentially retrograde rather then prograde. In this
paper we probe this conjecture by direct determination of Rheinland's
pole orientation (\S~2 and 3). Using this information, we revisit determination of
the age for this pair by backward tracking of its components' orbits
into the past (\S~4).

\section{Observations}\label{obs}
Previous photometry of Rheinland, from its favorable opposition in
2009, has been reported in the Supplementary materials of Pravec 
et~al. (2010). In this paper, we report additional observations from
three oppositions in 2008, 2009 and 2010-2011. Altogether
we thus present 34 lightcurves whose observation details, such
as the aspect data, heliocentric and observer distances, and observing
stations are given in Table~\ref{tab1}. A more detailed information
about the telescopes and data reduction procedures could be found
in the Supplementary materials of Pravec et~al. (2010).

The data from 2008 are limited, yet they are important for our modeling 
because they offer a new viewing geometry and help constraining the
precise value of the
sidereal rotation period. The data from the 2009 opposition are very
numerous, reach up to $28^\circ$ phase angles before and after
opposition and cover an interval of 4 months. This is because during the
opposition in September 2009 the asteroid was close to perihelion of its 
orbit and thus was quite bright, up to magnitude $15$ in
visible band. The data from 2010-2011 opposition are fewer, because
of fainter brightness, still they cover an interval of nearly 4
months too. They are less symmetrically distributed about the
opposition in March~2011, with less observations before and more
observation after the opposition. The sufficiently long periods
of time covered by observations in 2009 and 2010-2011 allow an 
unambiguous link of the data and provide a unique solution for the 
rotation period. During the 2009 and 2010-2011 oppositions, the 
geocentric ecliptic latitudes of the asteroid were different which 
suitably provides complementary aspects of view. However, due to a
small inclination of Rheinland's orbit with respect to the ecliptic, 
this latitude difference of the observations was still rather small.
As a result, determination of Rheinland's rotation pole longitude is
problematic and has larger uncertainty (\S~3).

Most of the data are on relative magnitude scales, either in clear or
R filters, but the three nights taken from Ond\v{r}ejov in February and
March 2011 were absolutely calibrated in the Cousins R system using Landolt 
standard stars. Using the parametrization of the phase function as 
described by the H-G system (e.g., Bowell et al. 1989), we derived
the best fit values for the absolute R magnitude $H_R = 13.68$ and the slope
parameter $G = 0.31$. Their formal errors, estimated accounting for 
uncertainties of the absolute calibrations, are $0.02$ and $0.03$, 
respectively. A systematic error of absolute magnitude estimated using 
the H-G function can be $\sim 0.05$~mag (see Harris 1991); we adopt 
this larger uncertainty for our estimated $H_R$. The absolute magnitude $H_R$
is that of the mean value over the lightcurve cycle.  Assuming
$V - R = 0.49\pm 0.05$, which is the mean color index for S-type 
asteroids (e.g., Shevchenko \& Lupishko 1998) that predominate in the 
inner main belt where Rheinland is located, we estimated its absolute 
$V$ magnitude $H = 14.17\pm 0.07$. Interestingly, this value is significantly
larger than $H = 13.6$ given by the MPC database or $H = 13.7$ given 
by the AstDyS databse (both use data from astrometric surveys). 
This example shows importance of the dedicated and accurate photometry 
in specific projects like analysis of the asteroid pairs.

\section{Pole and Shape of Rheinland}\label{pole}
We used the lightcurve inversion method of Kaasalainen \& Torppa (2001)
and Kaasalainen et~al. (2001) to derive Rheinland's shape, sidereal
rotation period and spin axis direction from the available data described
in \S~2.%
\footnote{The whole dataset of observations, parameters of the shape model and
 further information is available from the {\tt DAMIT} database at
 {\tt http://astro.troja.\lb{2}mff.cuni.cz/\lb{2}projects/\lb{2}asteroids3D/\lb{2}web.php} 
 (see also \v{D}urech et~al. 2010).}
We assume the body rotates about the shortest axis of the
inertia tensor which is fixed in the inertial space. This is because
(i) Fourier analysis of the individual lightcuves from different nights were
sufficiently well fitted with a single rotation period and its overtones
(due to irregular shape),%
\footnote{In fact, our observations confirm that (6070) Rheinland is
 very close to the principal axis rotation mode, which is by itself an
 interesting result. Note that a characteristic timescale to damp a tumbling
 state is about $1$~My for this body (see, e.g., Harris 1994), while
 the age of the Rheinland-2001~NQ8 pair is much younger (\S~4). This
 implies that the disruption process that has led to this pair formation
 was very gentle and did not excite Rheinland's rotation. Actually, the
 same conclusion holds also for many primaries in sub-My old pairs
 analysed by Pravec et~al. (2010).}
and (ii) gravitational and radiative torques
can change the spin state only on much longer timespan than the 4 years
between the first and the last observations. The Fourier fits of the
individual lightcurves were also used to estimate statistical
uncertainty of the individual measurements, a task which is characteristically
murky for the asteroid photometry. This is because number of systematic
sources of errors may prevent assignment of a clean, Gaussian-type
uncertainty to the measurements. Still, we are thus able to discriminate
between data with a very low scatter of the neighboring measurements from
those with large scatter of the neighboring measurements, and assign
appropriate relative weights to the data. We also assume a convex
shape represented with a polyhedron of a certain number (typically hundreds 
to thousands) of surface facets whose areas are given by the exponential
representation described in Kaasalainen \& Torppa (2001). We only
consider a combination of the Lommel-Seeliger and Lambert scattering of 
the sunlight on the surface of the asteroid. The method seeks to adjust
free parameters in order to minimize a target function of a $\chi^2$-type.%
\footnote{We have $\chi^2=\frac{1}{N-M}\sum_{i=1}^N (O-C)_i^2/\sigma_i^2$, where
 $N$ is the total number of observations and $M$ is the number of solved-for
 parameters of the model, $\sigma_i$ their estimated
 uncertainty from the analysis of observations scatter about Fourier
 representation of the individual lightcurves, and $(O-C)_i$ is the difference
 between the observed and computed brightness. For relative photometry, the
 lightcurves can be arbitrarily shifted on the magnitude scale.}

\begin{figure}
\epsscale{1.0}
%%% \hspace*{2.5cm}
 \plotone{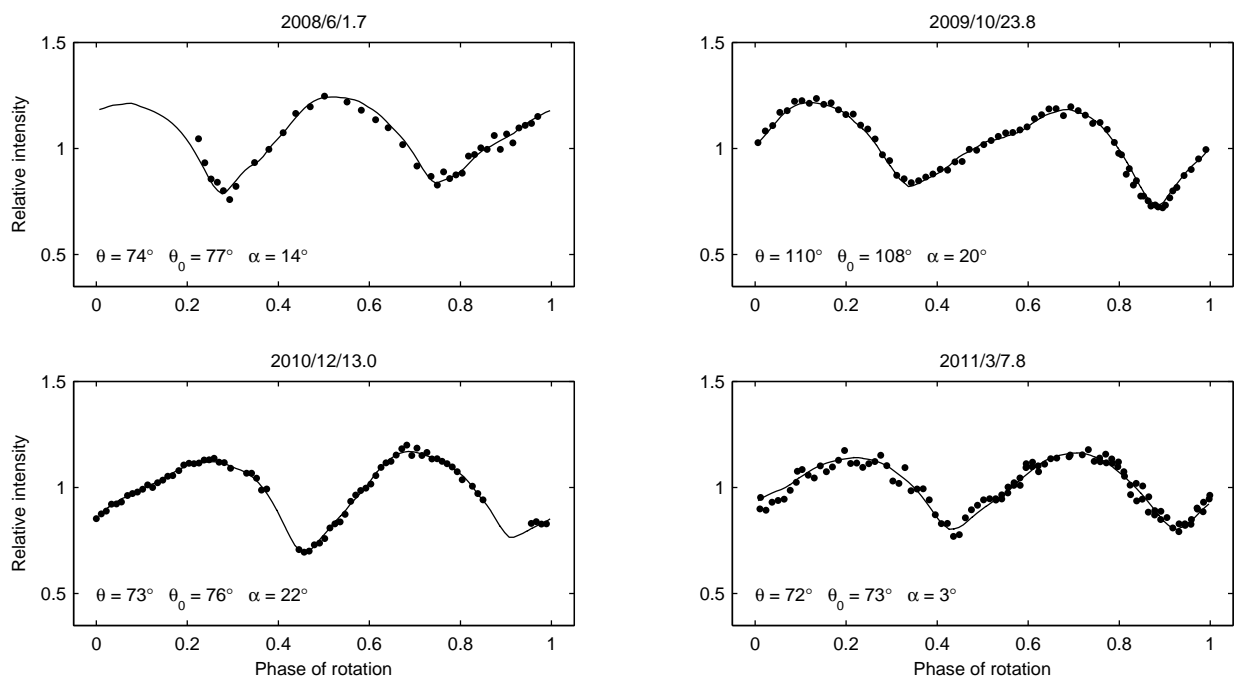}
\caption{Sample of Rheinland's photometric data (symbols) fitted
 with synthetic lightcurves based on the convex shape model (solid line).
 We used the formally best-fit model with pole orientation $(\lambda,\beta)=
 (4^\circ,-76^\circ)$ in ecliptic longitude and latitude, and sidereal rotation
 period $P=4.27371$~hr. The viewing and illumination geometry is given by the
 aspect angle $\theta$, the solar aspect angle $\theta_0$, and the solar phase
 angle $\alpha$.}
 \label{lcs}
\end{figure}

\begin{figure}
\epsscale{1.0}
%%% \hspace*{2.5cm}
 \plotone{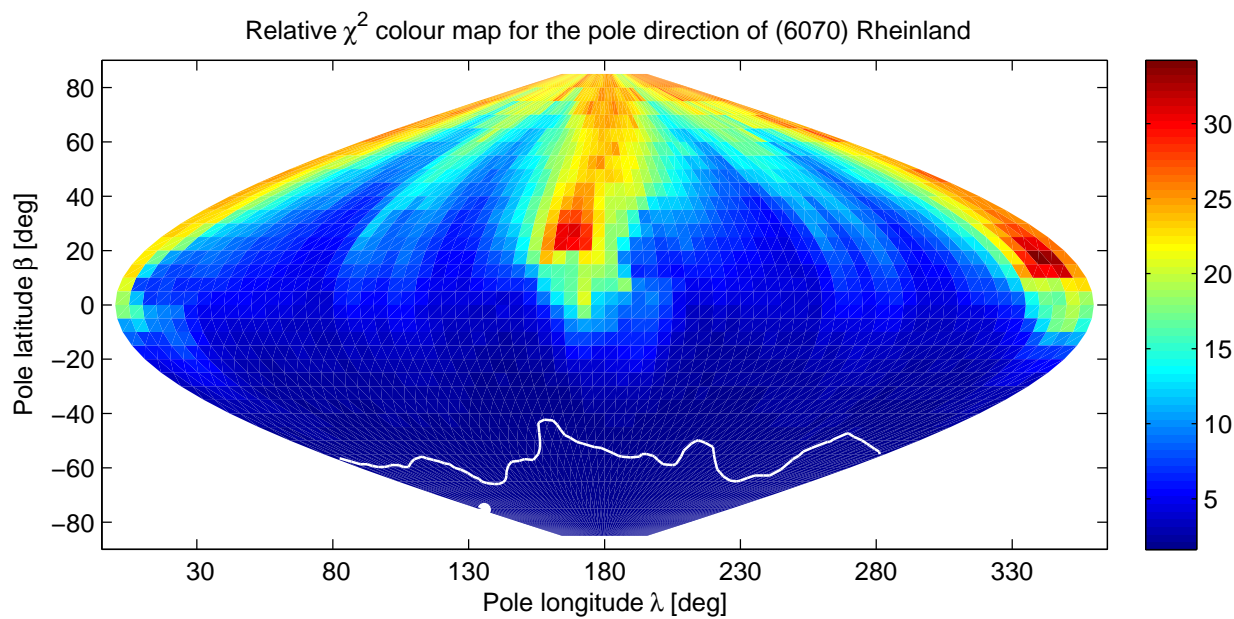}
\caption{Statistical quality of Rheinland's pole solutions shown in a
 sinusoidal projection of the sky in ecliptic coordinates. The grade of
 shading, and the scale bar on the right, indicates the value of $\chi^2$
 value normalized by the number of observations. The globally best-fit
 solution at $(\lambda,\beta)=(4^\circ,-76^\circ)$ (full circle) has
 $\chi^2=1.6$. The solid line, delimits solutions with 10\% larger $\chi^2$
 value than the best-fit solution, represents our region of admissible
 solutions (see the main text for details).}
 \label{polesky}
\end{figure}

Our best-fit solution has sidereal rotation period $P=4.27371$~hr and
rotation pole at $(\lambda,\beta)=(4^\circ,-76^\circ)$, where
$\lambda$ and $\beta$ are ecliptic longitude and latitude. Figure~\ref{lcs}
shows a sample of lightcurve data compared to the model. The pole position
is, however, not strongly constrained and using the longitude-latitude
parametrization we cannot simply assign some formal uncertainties.
Rather, we show in Figure~\ref{polesky} a whole-sky map of the $\chi^2$
values for individually best-fitted shape models. Since the $\chi^2$
values were normalized by number of degrees of freedom, the solutions
with $\chi^2\simeq 1$ would formally match the data in a statistical
sense. However, we recall that the photometric observation uncertainties
may not strictly-speaking obey the Gaussian statistics and that also
systematic and modeling errors are important. For these reasons, the globally
best-fit solution has $\chi^2=1.6$. To make the best-fit solution 
statistically acceptable, we would have to increase the formal errors 
of the measurements by about 25\%. The $\chi^2$-isocontour
shown in Figure~\ref{polesky} corresponds to solutions with 10\% larger
$\chi^2$ value than the global minimum (i.e., $\chi^2\simeq 1.8$),
which we consider still admissible. Because in our case $N\simeq 1750$ 
and the number of parameters $M\simeq 100$, the number of degrees of
freedom is $\nu = N - M \simeq 1650$, and the 10\% increase of $\chi^2$
corresponds to about $3\sigma$ interval of the $\chi^2$ distribution with
$\nu$ degrees of freedom.%
\footnote{The $\chi^2$ distribution with $\nu$ degrees of freedom has mean
 $\nu$ and variance $2\nu$ (e.g., Press et~al. 2007).}
We consider solutions with $\chi^2> 1.8$ values to be inadmissible,
as, indeed, they show too large inconsistencies between the observed and
computed magnitudes. Adopting this approach to the estimation of the 
uncertainty of our model, we may conclude that the ecliptic longitude of 
Rheinland's pole is not well
constrained yet, but the ecliptic latitude must be smaller than
$\simeq -50^\circ$. With only very small inclination of the orbit
with respect to the ecliptic plane (its proper value is $\sim
2.18^\circ$), our result thus implies that the obliquity $\varepsilon$
of Rheinland's pole is $\geq 140^\circ$, with the best-fit solution
value of $\simeq 165^\circ$. Rotation period solutions of Rheinland within 
the admissible zone differ by at most $\simeq 2\times 10^{-5}$~hr. We
can thus consider this value as a realistic uncertainty of our
solution for the sidereal rotation period for (6070)~Rheinland.

\begin{figure}
\epsscale{1.0}
%%% \hspace*{2.5cm}
 \plotone{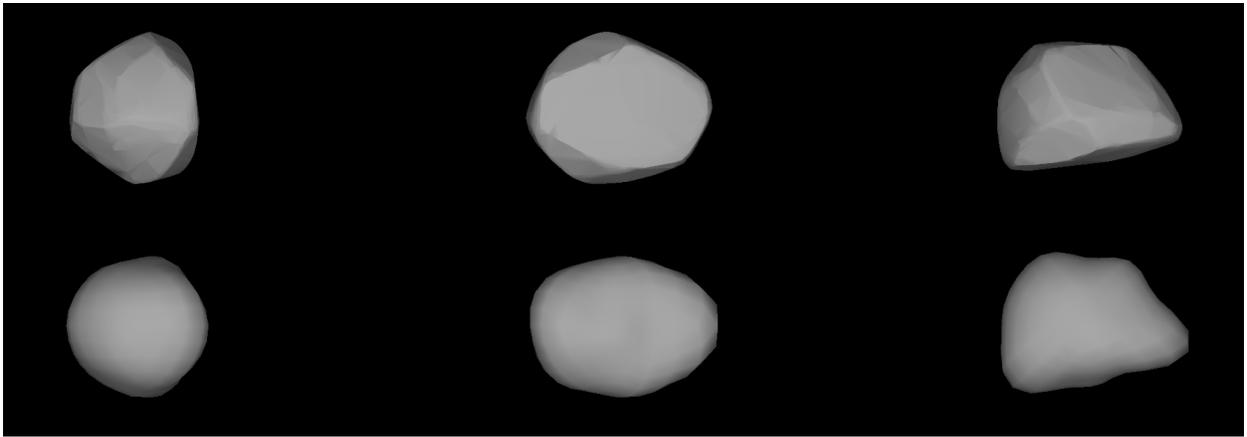}
\caption{Shape model for (6070) Rheinland from the lightcurve inversion analysis.
 We show two variants of the formally best-fit model: (i) a convex model in the
 top panels, and (ii) a non-convex model in the bottom panels. This latter is,
 however, not unique, and we give it as an example only. The three views are
 from the equatorial level (left and center) and  the pole-on (right).}
 \label{shape}
\end{figure}

The best-fit shape of Rheinland is shown in Figure~\ref{shape}. The
convex representation with 2038 surface facets is shown at the top
panels. Panels at bottom show, for sake of interest, a non-convex model
which has basically the same $\chi^2$ value as the convex-shape solution. In
general, the photometry of main belt asteroids, such as Rheinland,
cannot unambiguously reveal non-convex features of their surface
(e.g., \v{D}urech \& Kaasalainen 2003). The leftmost and rightmost
views on Figure~\ref{shape} indicate that our shape model of Rheinland
has a sharp, planar-like edge. While the lightcurve dataset is still
not very abundant, and our shape modeling may thus have its limitations, we 
note that this feature is correlated with the observed steep lightcurve 
decreases (see, e.g., near the phase 0.8 at the right and
top panel on Figure~\ref{lcs}) and cannot be entirely artificial. It is
tempting to hypothesize, that this feature may correspond to the
surface zone where the secondary component 2001~NQ8 separated from
the parent body of this pair. Further photometric observations of
Rheinland are important not only to shrink the persisting uncertainty
in the pole position, but also to confirm this interesting
surface feature. Unfortunately, the next favorable opposition which
will provide novel viewing geometry on the asteroid, and the target
will be bright enough, starts only in November 2013 and lasts till
January 2014.

\section{Implications and Discussion}\label{impl}
The above obtained constraint of the pole orientation for (6070) Rheinland 
may help us to refine determination of its age using backward integration of
orbits of the two components in this pair. This is because the
known obliquity importantly constrains the value of Yarkovsky
effect, one of the two factors that limit our ability of an accurate
(deterministic) orbital reconstruction in the past.

\subsection{Backward Orbital Integrations}
Detailed description of the age determination of a given pair of
asteroids using backward integration of their orbits was given by
Vokrouhlick\'y \& Nesvorn\'y (2008, 2009). Here we only outline
the main features of the approach, especially if relevant to 
findings in this paper.

The currently best-fit osculating orbits of both (6070) Rheinland (primary)
and (54827) 2001~NQ8 (secondary), derived from the 
available astrometric observations, are given in Table~\ref{tab2}.
These data were taken from {\tt AstDyS} database provided by 
University of Pisa (see \httppisa). Both orbits are fairly
well constrained at a comparable level, reflecting that both asteroids
have been observed over many oppositions and hundreds of astrometric
measurements are available for each of them. Table~\ref{tab2} gives
information about the uncertainty of the six orbital osculating elements 
${\bf E}$, but the complete solution obviously provides also the full 
covariance matrix $\mbox{\boldmath$\Sigma$}$ of the orbital fit, from 
which mutual correlations can be derived. While these correlations are
only moderately significant, with the largest correlation of $\sim 80$\%
between the semimajor axis and longitude in orbit solutions, it is
important to take them into account. Based on this information, we 
construct probability density distribution $p({\bf E})\propto 
\exp\left[-\frac{1}{2} \Delta{\bf E}\cdot \mbox{\boldmath$\Sigma$}
\cdot \Delta{\bf E}\right]$ (e.g., Milani \& Gronchi 2010), where
$\Delta{\bf E}={\bf E}-{\bf E}^\star$ with ${\bf E}^\star$ the 
best-fit orbital values given in Table~\ref{tab2}. All solutions 
${\bf E}$ with high-enough value of $p({\bf E})\geq C$, where $C$ is
related to a given confidence level, are statistically
equivalent and thus we cannot consider ${\bf E}^\star$ as the
only orbital realization of either primary or secondary components
in our pair of asteroids. Choosing a number of orbits which
will represent each of the asteroids in our numerical simulation, 
we used $p({\bf E})$ to determine their initial orbital values 
${\bf E}$. We call these different initial orbital realizations
geometrical clones. The geometrical clones occupy a six-dimensional
ellipsoid in the ${\bf E}$-space, or --after an appropriate 
transformation-- a six-dimensional ellipsoid region in the Cartesian
space of heliocentric positions and velocities.
When restricted to a three-dimensional space of heliocentric
positions, the geometrical clones occupy a three-dimensional ellipsoid
region with the longest axis approximately 200~km and 400~km for the
primary and secondary components in the Rheinland-2001~NQ8 pair. This 
shows how tightly constrained are both orbits at the initial epoch. For sake
of comparison with our convergence efforts described below, we note 
that the size of both uncertainty ellipsoids today is smaller than
the radius of the Hill sphere of influence of the primary 
(Rheinland) component (approximately 1000~km).

When propagated backward in time, the region occupied by geometric
clones extends. In the case of Rheinland-2001~NQ8 pair, and over
the relevant $\simeq 17.2$~kyr timescale of its age, this extension
is basically a simple stretching in the along-track direction by the
Keplerian shear of initial orbits with slightly different values
of the semimajor axis (in angular terms, the uncertainty translates to
about $\pm 0.02^\circ$ uncertainty in the longitude of orbit). 
This is because the orbits are not affected by
any of the major resonances. So while the short axes of the
uncertainty ellipsoid only slightly increase with respect to its
initial size, the long axis stretches to about $3\times 10^5$~km
some $\simeq 17.2$~kyr ago. This represents about 300 times the
radius of the Hill sphere of influence of Rheinland.

This uncertainty is very small and would have allowed even more
precise age determination of the pair if there were not for the second
source of the uncertainty in the past ephemerides for both components.
This latter effect is due to uncertainty in the dynamical model, in
particular parameters that influence strength and direction of the
thermal accelerations known as the Yarkovsky effect (Bottke et~al.
2002, 2006). The main orbital perturbation by the Yarkovsky effect is
a secular change in the semimajor axis, whose magnitude and sign
depends on asteroid's size, surface thermal inertia and rotation
state. While the asteroid's size can be roughly estimated from the
absolute magnitude and assumed value of geometric albedo, the surface
thermal inertia and rotation state are apriori unknown from astrometric
observations. Thermal inertia influences only the magnitude of
the effect to a factor which is typically not more than $\sim 5$
(e.g., Vokrouhlick\'y et~al. 2000), however the spin axis obliquity
value determines the overall sign of the semimajor axis drift:
for prograde-rotating asteroids the semimajor axis increases in time 
while for the retrograde-rotating asteroids decreases in time.
As a result, having been able to constrain Rheinland's obliquity
value, we remove a significant degree in uncertainty in its
past orbital evolution. As described in Vokrouhlick\'y et~al. (2000),
the semimajor axis secular change due to the Yarkovsky effect
$da/dt$ directly propagates into a quadratic perturbation in the
longitude in orbit. The Yarkovsky effect thus adds additional
component to the orbital stretching in the long-track 
direction, and over the $\simeq 17.2$~kyr timescale it
becomes more important than the effect of the initial orbit uncertainty.
Using Eq.~(30) in Vokrouhlick\'y et~al. (2000), we obtain $\pm (0.6^\circ
-0.7^\circ)$ longitude in orbit uncertainty of the Rheinland's orbit $\simeq
17.2$~kyr ago.%
\footnote{We used $\sim 3.9$~km size estimate from the absolute magnitude
 determined in \S~2 and assumed geometric albedo $\simeq 0.25$, and thermal
 inertia $\sim 200$~J/m$^2$/s$^{0.5}$/K, appropriate mean value for small
 asteroids of Rheinland's size (e.g., Delb\`o et~al. 2007). Since we consider
 an albedo value near the upper limit of the S-type class asteroids of
 Rheinland's absolute magnitude, the obtained size is rather an
 underestimate. As in Vokrouhlick\'y \& Nesvorn\'y (2008, 2009) we adopt this
 conservative approach not to exclude any possible Yarkovsky drift-rates
 of the semimajor axis from our analysis.}
This is $\simeq 30$ times more than the spread of geometrical clones
at the same time. Because the Yarkovsky effect magnitude is indirectly
proportional to the asteroid's size, the along-track uncertainty
is even larger for the secondary component (54827) 2001~NQ8, for which
in amounts to $\pm (1.1^\circ-1.3^\circ)$ longitude in orbit uncertainty.
This is again an effect $\simeq 30$ times larger than that produced by
the uncertainty of the initial orbital data for this asteroid.

We model the influence of unconstrained Yarkovsky effect by assigning to each
geometric clone a spectrum of Yarkovsky accelerations. We call these
different orbital variants Yarkovsky clones. To simplify computations, we
represent the Yarkovsky acceleration by an empirical transverse acceleration
with a magnitude determined by the modeled rate $da/dt$ of the secular
change in semimajor axis (e.g., Vokrouhlick\'y \& Nesvorn\'y 2008, 2009).
We used {\tt SWIFT\_MVS} numerical integrator for orbit propagation to
the past (e.g., Levison \& Duncan 1997) with a fixed timestep of $5$~days.
Perturbations due to all planets, whose initial data at MJD 55600 were
taken from the JPL DE405 ephemerides, are included. The empirical
formulation of the thermal forces, as described above, has been added
to the code. We propagated 20 geometrical and 30 Yarkovsky clones for both
primary and secondary components in the Rheinland-2001~NQ8 pair, so
altogether 600 clones for each asteroid, and examined online their mutual
distances every $0.25$~y during the orbital propagation. Because we
had a preliminary knowledge of the age for this pair, we integrated
orbits of all clones to $20$~kyr in the past. Velocity components of
the initial data, both planets and asteroid clones, were reversed and
integration timestep was positive. With that setting, the true drift 
rate values of the semimajor axis are reversed. So, while the obliquity
$\geq 140^\circ$ for (6070) Rheinland implies negative value of $da/dt$,
we assigned formally positive $da/dt$ value to the Yarkovsky clones of this
asteroid in our backward integration (to prevent confusions, though, we
use the true $da/dt$ values in what follows). Because of the unknown value
of the surface thermal inertia of Rheinland, we conservatively considered
all values $da/dt$ between $-5.3\times 10^{-5}$~AU/My and $0$
(appropriate for this asteroid size, Bottke et~al. 2002, 2006). In the 
case of the secondary component, (54827) 2001~NQ8, we took Yarkovsky 
clones with both positive and negative $da/dt$ values. For sake of a 
more detailed analysis below, we actually ran two simulations, first 
with 30 clones of 2001~NQ8 and $da/dt$ positive and second with 30 
clones of 2001~NQ8 and $da/dt$ negative. The maximum $|da/dt|$ value 
in this case was $10^{-4}$~AU/My, because the secondary component in
the pair has about half the size of the primary.

As described above, some $17$~kyr ago the regions of uncertainty in the
past ephemerides occupied by the geometric and Yarkovsky clones of both 
components the Rheinland-2001~NQ8 pair resemble very elongated ellipsoids 
in Cartesian space. Their long axes are $\sim 4000$ times for Rheinland, resp.
$\sim 15000$ times for 2001~NQ8, the estimated Hill sphere of influence of Rheinland
which is the quantitative measure of the orbital convergence (see, e.g.,
Vokrouhlick\'y \& Nesvorn\'y 2009). Henceforth only a fraction of
propagated clones result in a successful convergence in our
numerical experiment. In practice, every $0.25$~y step in our propagation 
we compute relative distance and velocity of each Rheinland's clone and 
each 2001~NQ8's clone. We consider the configuration to be convergent
when the clone distance is less than 75\% of the instantaneous Hill sphere of 
Rheinland (typically $\simeq 750$~km) and their relative velocity is
less than $\simeq 2$~m/s (i.e., the estimated escape velocity from
Rheinland). Examining these convergent cases not only provides a
constraint of the age for this pair, but it may also provide additional
information such as preference between the Yarkovsky clones of the
secondary component, 2001~NQ8, with positive or negative $da/dt$
values. 

\begin{figure}
\epsscale{0.8}
%%% \hspace*{2.5cm}
 \plotone{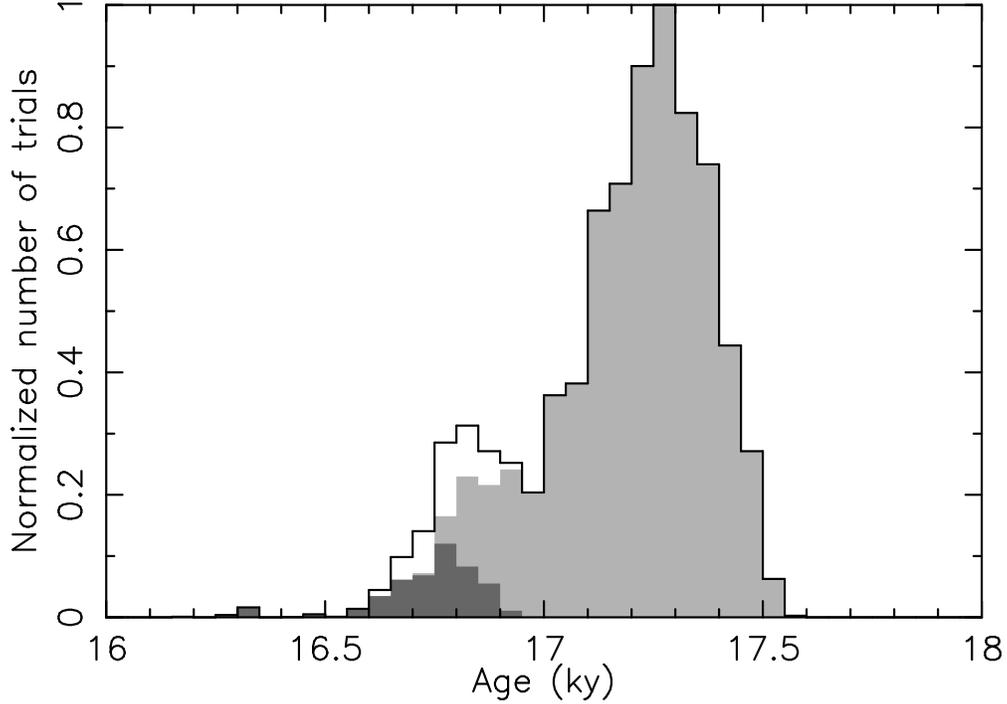}
\caption{Distribution of number of trials that resulted in a satisfactory
 convergence solution of (6070) Rheinland and (54827) 2001 NQ8 orbits:
 time in the past at the abscissa (in ky) and a normalized histogram of
 converging solutions in 50~yr bins at the ordinate. We used pairs of clones that
 approached closer than $750$~km and had relative velocity smaller than 
 $2$~m/s. All clones of (6070) Rheinland had negative secular drift
 in the semimajor axis due to the Yarkovsky effect in agreement with the
 pole solution from \S~3. The clones of (54827) 2001 NQ8 had both positive
 and negative drift in the semimajor axis. The open histogram shown by the
 enclosing solid line corresponds to all cases; its maximum also serves for
 the normalization. The light-gray histogram corresponds to the cases where
 the (54827) 2001 NQ8 clones had negative drift in the semimajor axis, while
 the dark-gray histogram corresponds to the cases where the (54827) 2001 NQ8
 clones had positive drift in the semimajor axis. The mean value and formal
 standard deviation of the distributions are $17.2\pm 0.2$~kyr in the first
 case and $16.75\pm 0.15$~kyr in the second case (the realistic uncertainty might
 be slightly larger due to a non-Gaussian nature of distribution functions).}
 \label{sol}
\end{figure}

Figure~\ref{sol} shows results of our backward tracking of clones for
both primary and secondary components in the Rheinland-2001~NQ8 pair.
The light gray histogram corresponds to the run where the Yarkovsky
clones of 2001~NQ8 had $da/dt<0$, i.e. the same sign as those of
Rheinland. The black histogram corresponds to the run where the Yarkovsky
clones of 2001~NQ8 had $da/dt>0$, i.e. the opposite sign as those of
Rheinland. The former case thus means the rotation of 2001~NQ8 has
same (retrograde) sense as that of Rheinland, while the latter case
implies the opposite. There are about 15 times more successful convergence
solutions in the former case than in the latter. The mean and standard
deviation values of the age estimates are $17.2\pm 0.2$~kyr for the
former case and $16.75\pm 0.15$~kyr in the latter case. As it has been
suggested above, not all combinations of clones provide convergent
configurations: at best, we had $\sim 10^{-3}$ fraction success. Since the
long axes of the ellipsoids occupied by clones have been estimated
to $\sim 4000$, resp. $\sim 7500$, Hill spheres of Rheinland,%
\footnote{The smaller value of the long axis for clones of 2001~NQ8,
 as compared to that given above, is because we propagate cases for
 positive and negative Yarkovsky drift rates in two different simulations.}
while the short axes are comparable to the Hill sphere of Rheinland, the
$\sim 10^{-3}$ success rate for convergence implies a very small tilt
between the long axes of the uncertainty ellipsoids of Rheinland and
2001~NQ8 clones. Indeed, our convergent solutions were
always characterized with a very small relative velocity of the order
$10-30$~cm/s, implying very similar orbits (see also Vokrouhlick\'y \&
Nesvorn\'y 2008, 2009).

\subsection{Rotation State of (54827) 2001~NQ8 and Formation Scenario}
While we obtained some convergent solutions for the opposite rotation
sense of the secondary component 2001~NQ8 as compared to Rheinland,
we had an order-of-magnitude more solutions for the same
sense of rotation of both components in the pair. If we were to attribute
a pure statistical meaning to this difference, we were to conclude 
that the case of parallel spin orientations of both components in 
the Rheinland-2001~NQ8 pair is a more likely case. Obviously,
such a conclusion is problematic because so far we do obtain
convergence solution for both spin orientations of 2001~NQ8. It thus
appears that determination of the rotation state for 2001~NQ8 is
the key element for both better determination of this pair's age but 
also for constraining the formation process. Note, for instance,
that the $\simeq 5.8764$~hr rotation period of the secondary by
itself favors ``a prompt ejection scenario'' as opposed to 
``a destabilization of a binary scenario'' (see Pravec et~al. 2010),
but knowing the pole orientation of 2001~NQ8 would provide a much
more complete information. The analysis would be eased if it were 
indeed near-to-parallel with the pole of Rheinland, as hinted here,
because the spin-orbit secular resonances do not affect the 
retrograde rotation states (e.g., Vokrouhlick\'y et~al. 2006).

\subsection{Future Fate of (6070) Rheinland}
While the solution of the rotation state and shape of Rheinland in
\S~3 is still very limited, we may use it to estimate the value of 
a secular change in its rotation rate $\upsilon=d\omega/dt$ due to
the YORP effect. One should take this exercise as an example of
interest rather than a true prediction, since the YORP effect
have been shown to eventually depend on many unknown or
inaccurately known parameters such as the small-scale structures
of the asteroid shape (e.g., Statler 2009; Breiter et~al. 2009) or
inhomogeneities in the density distribution (e.g., Scheeres \& Gaskell
2008). Taking thus the best fit solution for Rheinland's shape and
rotation state from \S~3 we obtain $\upsilon\simeq 10^{-9}$~rad/d$^2$.
In terms of magnitude, this is about the expected value for an
asteroid of its size and heliocentric distance if we appropriately 
scale the directly detected YORP values for (54509) YORP (e.g., 
Lowry et~al. 2007; Taylor et~al. 2007), (1862) Apollo (e.g., 
Kaasalainen et~al. 2007) or (1620) Geographos (e.g., \v{D}urech et~al. 
2008). The positive sign of $\upsilon$ implies the rotation rate 
of Rheinland is accelerated by the YORP torques,
and in $\sim (50-100)$~My it may bring its rotation state to the
fission limit. Assuming the Rheinland-2001~NQ8 pair was actually born by
rotational fission of a precursor asteroid, this would have been
at least the second such event for the same body. While future
improved shape solutions for Rheinland, from larger observation datasets,
may modify our result, we consider this to be an example of a process
that may actually be frequent for small asteroids in the main
belt: a sequence of fission events driven by YORP torques that
continually erode the body by mass shedding and producing either 
paired secondaries or binary systems. We note that the estimated 
timescale above is quite shorter than the collisional lifetime of
Rheinland, some $\sim 1$~Gy according to Bottke et~al. (2005).
Unfortunately, the small value of $\upsilon$ means that we will not
be able to directly measure YORP effect for this asteroid any soon.%
\footnote{On the other hand, the long timescale for YORP acceleration
 of Rheinland's rotation rate is fortunate, since the currently
 observed rotation rate of Rheinland is the same as at the moment of
 separation. So the observed rotation periods of primaries of the
 pairs directly probe the separation process of components in the
 asteroid pairs (Pravec et~al. 2010).}
One can easily estimate that at least four to five decades with
suitably distributed data are necessary for this task.

\acknowledgements
 We thank Alan Harris for his thorough review which improved this paper.
 This research was supported by Czech Grant Agency (grants 205/08/0064 and 
 205/09/1107) and the Research Program MSM0021620860 of the Czech Ministry
 of Education. The work at Modra was supported by the Slovak Grant Agency
 for Science VEGA, Grant 2/0016/09. We thank A.~Marshalkina, M.~Ibrakhimov
 and A.~Sergeev for assistance in 2008 and 2010 observations at Maidanak
 Observatory.

\clearpage

% TABLE 1 -------------------------------------------------------------
\begin{deluxetable}{cccrrrl}
%\tabletypesize{\scriptsize}
\tabletypesize{\footnotesize}
\tablecaption{Aspect data for observations of (6070) Rheinland
 \label{tab1}}
\tablewidth{0pt}
\tablehead{\colhead{Date} & \colhead{$r$}  & \colhead{$\Delta$} & \colhead{$\alpha\phantom{g}$}
 & \colhead{$\lambda$} & \colhead{$\beta$} & \colhead{Obs.} \\
 \colhead{} & \colhead{(AU)} & \colhead{(AU)} & \colhead{(deg)} & \colhead{(deg)} &
 \colhead{(deg)} & \colhead{}}
\startdata
2008 05 14.9  & 2.826 & 1.860 &  7.5 & 213.2 & $ 3.3$ & Kh   \\
2008 05 31.7  & 2.809 & 1.958 & 13.5 & 210.3 & $ 2.9$ & Ma   \\
2008 06 01.7  & 2.808 & 1.965 & 13.8 & 210.2 & $ 2.9$ & Ma   \\
2009 07 22.0  & 1.981 & 1.267 & 26.4 & 359.1 & $-4.6$ & W1   \\
2009 07 24.0  & 1.978 & 1.246 & 25.9 & 359.4 & $-4.7$ & W1   \\
2009 07 25.0  & 1.976 & 1.237 & 25.7 & 359.6 & $-4.7$ & W1   \\
2009 07 26.0  & 1.975 & 1.226 & 25.5 & 359.8 & $-4.7$ & W1   \\
2009 07 27.0  & 1.973 & 1.217 & 25.3 & 359.9 & $-4.8$ & Kh   \\
2009 07 28.0  & 1.972 & 1.208 & 25.0 &   0.0 & $-4.8$ & Kh   \\
2009 07 32.0  & 1.965 & 1.170 & 24.0 &   0.5 & $-5.0$ & Kh   \\
2009 08 01.9  & 1.964 & 1.161 & 23.7 &   0.6 & $-5.1$ & Kh   \\
2009 08 17.9  & 1.941 & 1.032 & 18.1 &   1.1 & $-5.8$ & W1   \\
2009 08 19.0  & 1.940 & 1.024 & 17.6 &   1.0 & $-5.8$ & Kh   \\
2009 09 20.9  & 1.905 & 0.904 &  3.7 & 355.7 & $-6.6$ & W1   \\
2009 10 20.8  & 1.888 & 0.994 & 18.4 & 351.1 & $-5.7$ & W1   \\
2009 10 23.8  & 1.887 & 1.012 & 19.7 & 351.0 & $-5.6$ & W1   \\
2009 11 17.8  & 1.886 & 1.209 & 27.6 & 353.4 & $-4.4$ & Si   \\
2009 11 18.7  & 1.887 & 1.217 & 27.8 & 353.6 & $-4.3$ & Si   \\
2009 11 20.8  & 1.887 & 1.237 & 28.2 & 354.1 & $-4.2$ & W1   \\
2009 11 22.7  & 1.888 & 1.255 & 28.5 & 354.5 & $-4.1$ & W1   \\
2010 12 13.0  & 2.621 & 2.359 & 22.0 & 166.5 & $ 3.0$ & Ma   \\
2010 12 17.0  & 2.629 & 2.311 & 21.8 & 167.2 & $ 3.1$ & Ma   \\
2011 01 08.9  & 2.670 & 2.051 & 18.8 & 169.1 & $ 3.7$ & Ma   \\
2011 01 09.9  & 2.672 & 2.040 & 18.6 & 169.1 & $ 3.8$ & Ma   \\
2011 02 24.5  & 2.744 & 1.761 &  3.0 & 162.2 & $ 4.7$ & Oey  \\
2011 02 26.5  & 2.747 & 1.761 &  2.3 & 161.7 & $ 4.7$ & Oey  \\
2011 02 27.0  & 2.747 & 1.761 &  2.2 & 161.5 & $ 4.7$ & Ond  \\
2011 03 01.9  & 2.752 & 1.763 &  1.7 & 160.8 & $ 4.8$ & Mo   \\
2011 03 03.9  & 2.754 & 1.765 &  1.9 & 160.4 & $ 4.8$ & W1   \\
2011 03 07.8  & 2.759 & 1.774 &  3.1 & 159.4 & $ 4.8$ & W2   \\
2011 03 08.8  & 2.761 & 1.778 &  3.6 & 159.0 & $ 4.8$ & Ond  \\
2011 03 28.9  & 2.787 & 1.898 & 11.3 & 154.8 & $ 4.6$ & Ond  \\
2011 04 01.8  & 2.791 & 1.931 & 12.6 & 154.4 & $ 4.5$ & W1   \\
2011 04 02.0  & 2.792 & 1.932 & 12.6 & 154.4 & $ 4.5$ & W2   \\
\enddata

\tablecomments{\footnotesize The whole dataset of Rheinland's observations used
 in our work. All but those from Wise observatory in 2009, which were 
 already reported in the Supplementary materials of Pravec et~al. (2010),
 are new data. The table gives Rheinland's distance from the Sun $r$ and from
 the Earth $\Delta$, the solar phase angle $\alpha$, the geocentric ecliptic
 coordinates of the asteroid $(\lambda,\beta)$, and the observatory (Mo -- Modra
 Observatory, 0.6 m; Ma - Maidanak Observatory, 1.5 m; Oey - Leura Observatory, 0.35 m;
 W1 - Wise Observatory, 0.46 m; W2 - Wise Observatory, 1 m; Ond - Ond\v{r}ejov
 Observatory, 0.65 m; Kh - Kharkiv Observatory, 0.7 m; Si - Simeiz Observatory, 1 m).}

\end{deluxetable}

\clearpage

% TABLE 2 -------------------------------------------------------------
\begin{deluxetable}{rlccccccc}
%\tabletypesize{\scriptsize}
\tabletypesize{\footnotesize}
\tablecaption{Osculating orbital elements, their uncertainties and other
 parameters of the asteroid pair (6070)~Rheinland and (54827) 2001~NQ8.
 \label{tab2}}
\tablewidth{0pt}
\tablehead{
 \multicolumn{2}{c}{Asteroid} & \colhead{$a$} & \colhead{$h$} & \colhead{$k$} &
  \colhead{$p$} & \colhead{$q$} & \colhead{$\lambda$} & \colhead{$H$} \\
 \colhead{} & \colhead{} & \colhead{(AU)} & \colhead{} & \colhead{} &
 \colhead{} & \colhead{} & \colhead{(deg)} & \colhead{(mag)}}
\startdata
 6070 & Rheinland & 2.388143165 & 0.06019115 & 0.20141467 &
        0.02717789 &  0.00285789 & 138.859782 & 14.17 \\
54827 & 2001 NQ8  & 2.388531447 & 0.06005468 & 0.20149015 &
        0.02716790 &  0.00285259 & 174.303041 & 15.2\phantom{7} \\
[1.5ex]
 \multicolumn{9}{c}{{\em Uncertainty\, ($\delta a,\delta h,\delta k,\delta p,
                                         \delta q,\delta\lambda,\delta H$)}} \\
[1.5ex]
 6070 & Rheinland & 1.9e-8 & 7.0e-8 & 9.0e-8 & 6.7e-8 & 8.1e-8 & 1.0e-5 & 0.07 \\
54827 & 2001 NQ8  & 4.5e-8 & 1.0e-7 & 1.3e-7 & 9.0e-8 & 1.1e-7 & 1.7e-5 & 0.5? \\
\enddata

\tablecomments{\normalsize Osculating orbital elements and their uncertainty are
 given for epoch MJD 55600 provided by the {\tt OrbFit9} software (\httppisa).
 We use heliocentric equinoctical system of non-singular elements as of
 May~2011: $a$ is the semimajor axis, $(h,k)= e\,(\sin\varpi,\cos\varpi)$ where
 $e$ is the eccentricity and $\varpi$ is the longitude of perihelion, $(p,q)=
 \tan (i/2)\, (\sin\Omega,\cos\Omega)$ where $i$ is the inclination and $\Omega$
 is the longitude of node, and $\lambda=\varpi+M$ is the mean longitude in orbit
 ($M$ is the mean anomaly). Default reference system is that of mean ecliptic
 of J2000. In the case of the primary component, (6070) Rheinland, we use
 absolute magnitude $H$ value determined in \S~2. In the case of the secondary
 component, (54827) 2001~NQ8, we adopted the absolute magnitude $H$ value given
 by the Minor Planet Center.}

\end{deluxetable}

\clearpage

\clearpage

\clearpage

\clearpage

\begin{thebibliography}

\bibitem{betal02} Bottke, W. F., Vokrouhlick\'y, D., Rubincam, D. P., \&
 Bro\v{z}, M. 2002, in Asteroids III, ed. W. F. Bottke et~al. (Tucson:
 University of Arizona Press), 395

\bibitem{betal06} Bottke, W. F., Vokrouhlick\'y, D., Rubincam, D. P., \&
 Nesvorn\'y, D. 2006, Ann. Rev. Earth Planet. Sci., 34, 157 

\bibitem{betal05} Bottke, W. F., Durda, D. D., Nesvorn\'y, D., Jedicke, R.,
 Morbidelli, A., Vokrouhlick\'y, D., \& Levison, H. F. 2005, Icarus, 179, 63

\bibitem{betal89} Bowell, E., Hapke, B., Domingue, D., Lumme, K., Peltoniemi, J.,
 \& Harris, A. W., in Asteroids II, ed. M. S. Matthews et~al. (Tucson:
 University of Arizona Press), 524

\bibitem{betal09} Breiter, S., Bartczak, P., Czekaj, M., Oczujda, B., \&
 Vokrouhlick\'y, D. 2009, A\&A, 507, 1073  

\bibitem{detal07} Delb\`o, M., Dell'Oro, A., Harris, A. W., Mottola, S., \&
 Mueller, M. 2007, Icarus, 190, 236	

\bibitem{dk03} \v{D}urech, J., \& Kaasalainen, M. 2003, A\&A, 404, 709

\bibitem{detal0} \v{D}urech, J., et~al. 2008, A\&A, 489, L2
%\bibitem{detal0} \v{D}urech, J., \& 11 coauthors 2008, A\&A, 489, L25

\bibitem{detal10} \v{D}urech, J., Sidorin, V., \& Kaasalainen, M. 2010,
 A\&A, 513, A46

%\bibitem{ietal01} Ivezi\'c, Z., et~al. 2001, AJ, 122, 2749
%\bibitem{ietal01} Ivezi\'c, Z. \& 31 coauthors 2001, AJ, 122, 2749

\bibitem{h91} Harris, A. W. 1991, in Asteroids, Comets, Meteors 1991,
 ed. A. W. Harris \& E. Bowell, (Houston: Lunar and Planetary Institute), 85

\bibitem{h94} Harris, A. W. 1994, Icarus, 107, 209

\bibitem{kt01} Kaasalainen, M., \& Torppa, J. 2001, Icarus, 153, 24

\bibitem{ketal01} Kaasalainen, M., Torppa, J., \& Muinonen, K. 2001,
 Icarus, 153, 37

\bibitem{ketal07} Kaasalainen, M., \v{D}urech, J., Warner, B. D.,
 Krugly, Y. N., \& Gaftonyuk, N. M. 2007, Nature, 446, 420

\bibitem{ld94} Levison, H. F., \& Duncan, M. J. 1994, Icarus, 108, 18

\bibitem{letal07} Lowry, S. C., et~al. 2007, Science, 316, 272
%\bibitem{letal07} Lowry, S. C. \& 9 coauthors 2007, Science, 316, 272

\bibitem{mg10} Milani, A., \& Gronchi, G. F. 2010, Theory of Orbit
 Determination, Cambridge University Press, Cambridge

\bibitem{metal10} Milani, A., Kne\v{z}evi\'c, Z., Novakovi\'c, B., \&
 Cellino, A. 2010, Icarus, 207, 769

\bibitem{nv06} Nesvorn\'y, D., \& Vokrouhlick\'y, D. 2006, AJ, 132, 1950

\bibitem{netal06a} Nesvorn\'y, D., Vokrouhlick\'y, D., \& Bottke, W. F.
 2006, Science, 312, 1490

\bibitem{pv09} Pravec, P., \& Vokrouhlick\'y, D. 2009, Icarus, 204, 580

\bibitem{petal02} Pravec, P., Harris, A. W., \& Micha{\l}owski, T. 2002,
 in Asteroids III, ed. W. F. Bottke et~al. (Tucson: University of Arizona
 Press), 113

\bibitem{petal10} Pravec, P., et~al. 2010, Nature, 466, 1085
%\bibitem{petal10} Pravec, P. \& 25 coauthors 2010, Nature, 466, 1085

\bibitem{petal07} Press, W. H., Teukolsky, S. A., Vetterling, W. T., \&
 Flannery, B. P. 2007, Numerical Recipes (Cambridge University Press,
 Cambridge)

\bibitem{sg08} 	Scheeres, D. J., \& Gaskell, R. W. 2008, Icarus, 198,
 125

\bibitem{sl98} Shevchenko, V. G., \&  Lupishko, D. F. 1998, Solar System Res.,
 32, 220

\bibitem{s09} Statler, T. 2009, Icarus, 202, 502

\bibitem{tetal07} Taylor, P. A., et~al. 2007, Science, 316, 274
%\bibitem{tetal07} Taylor, P. A. \& 11 coauthors 2007, Science, 316, 274

\bibitem{vn08} Vokrouhlick{\'y}, D., \& Nesvorn\'y, D.\ 2008, AJ, 136, 280 

\bibitem{vn09} Vokrouhlick{\'y}, D., \& Nesvorn\'y, D.\ 2009, AJ, 137, 111 

\bibitem{vn11} Vokrouhlick{\'y}, D., \& Nesvorn\'y, D.\ 2011, AJ, 142, 26

\bibitem{vetal00} Vokrouhlick{\'y}, D., Milani, A., \& Chesley, S. R.\ 2000,
 Icarus, 148, 118

\bibitem{vetal06} Vokrouhlick{\'y}, D., Nesvorn\'y, D., \& Bottke, W. F.
 2006, Icarus, 184, 1

\end{thebibliography}
\end{document}